%% file: main_arxiv.tex
\documentclass[11pt]{article}

\usepackage[left=2.85cm,right=2.85cm,top=2cm,bottom=3cm]{geometry}

\usepackage{amsmath,amsfonts,amssymb}
\usepackage[toc,page]{appendix}
\usepackage[utf8]{inputenc}
\usepackage[T1]{fontenc}
\usepackage{graphicx}
\usepackage{epstopdf}
\usepackage{lscape}

\usepackage[font=small,labelfont=bf]{caption}
\usepackage{color}
\usepackage{customCommands}
\usepackage{rotating}
\usepackage{cleveref}
\crefname{figure}{Figure}{Figures}
\crefname{table}{Table}{Tables}
\crefname{section}{Section}{Sections}

\usepackage{multirow} 
\usepackage{color,soul}
\usepackage[]{footmisc}

\usepackage{enumerate}   
\usepackage{booktabs}
\usepackage{multicol}

\usepackage[boxed, ruled, lined]{algorithm2e}

\renewcommand{\argmin}{\operatorname*{argmin}}
\newcommand{\trace}{\operatorname{trace}}

\newcommand{\df}{\operatorname{DOF}}

\newcommand{\GCV}{\operatorname{GCV}}
\newcommand{\tBAT}{t_\text{BAT}}

\usepackage{enumerate}
\usepackage{csquotes}

\usepackage[authoryear,round,sort]{natbib}

\usepackage{url}

\title{Bolus arrival time estimation in dynamic contrast-enhanced magnetic resonance imaging of small animals based on spline models}

\author{Alina L. Bendinger$^{1,2}$,
Charlotte Debus$^{3,5,6}$,
Christin Glowa$^{4,5,6}$,\\
Christian P. Karger$^{4,6}$,
Jörg Peter$^1$,
Martin Storath$^{7,8}$
}

\date{\small
\begin{flushleft}
$^1$ Department of Medical Physics in Radiology, German Cancer Research Center (DKFZ), Heidelberg, Germany \\
$^2$ Faculty of Biosciences, University of Heidelberg, Heidelberg, Germany \\
$^3$ Translational Radiation Oncology, National Center for Tumor Diseases (NCT), German Cancer Research Center (DKFZ), Heidelberg, Germany \\
$^4$ Department of Medical Physics in Radiation Oncology, German Cancer Research Center (DKFZ), Heidelberg, Germany \\
$^5$ Department of Radiation Oncology and Radiotherapy, University Hospital Heidelberg, Heidelberg, Germany \\
$^6$ Heidelberg Institute for Radiation Oncology (HIRO) and National Center for Radiation Research in Oncology (NCRO), Heidelberg, Germany \\
$^7$ Image Analysis and Learning Group, Interdisciplinary Center for Scientific Computing, University of Heidelberg, 
Germany \\
$^8$ Faculty of Applied Natural Sciences and Humanities, University of Applied Sciences Würzburg-Schweinfurt, 
Germany
\end{flushleft}
}

\begin{document}

\maketitle
\begin{abstract}
Dynamic contrast-enhanced magnetic resonance imaging (DCE-MRI) is used to quantify perfusion and vascular permeability. In most cases a bolus arrival time (BAT) delay exists between the arterial input function (AIF) and the contrast agent arrival in the tissue of interest which needs to be estimated. Existing methods for BAT estimation are tailored to tissue concentration curves which have a fast upslope to the peak as frequently observed in patient data.
However, they may give poor results for curves that do not have this characteristic shape such as tissue concentration curves of small animals.
In this paper, we propose a novel method for BAT estimation of signals that do not have a fast upslope to their peak. The model is based on splines which are able to adapt to a large variety of concentration curves. Furthermore, the method estimates BATs on a continuous time scale. All relevant model parameters are automatically determined by generalized cross validation. We use simulated concentration curves of small animal and patient settings to assess the accuracy and robustness of our approach.
 The proposed method outperforms a state-of-the-art method for small animal data and it gives competitive results for patient data. 
Finally, it is tested on \textit{in vivo} acquired rat data where accuracy of BAT estimation was also improved upon the state-of-the-art method.
The results indicate that the proposed method is suitable for accurate BAT estimation of DCE-MRI data, especially for small animals.
\end{abstract}

\section{Introduction}

Dynamic contrast-enhanced magnetic resonance imaging (DCE-MRI) is used to study perfusion and vascular permeability in tissues
such as tumor vasculature \citep{sourbron2012pkm}. This makes DCE-MRI especially interesting for the characterization of the vascular status in clinical studies \citep{padhani2010imaging} and in preclinical studies employing small animals \citep{hectors2018dce, schreurs2017dce, glowa2017, keunen2011dce}.

DCE-MRI is based on acquiring MR-images at high temporal resolution after intravenously administering a contrast agent (CA) bolus. Concentration time curves describing the CA accumulation are extracted from the tissue of interest. Quantitative analysis of these tissue concentration curves (TCCs) is often performed by fitting pharmacokinetic models to them \citep{sourbron2012pkm}. Many of these models require an arterial input function (AIF) describing the arrival of the CA in an artery close to the tissue of interest \citep{sourbron2010perfusion, tofts1999etm}.
A common non-invasive technique for patient individual AIF extraction is to derive it directly from the acquired DCE-MRI images from a vessel close to the tissue of interest. Due to limited image resolution and partial volume effects it is not always possible to choose an artery in the direct vicinity of the tissue. Hence, the AIF is often extracted from a larger artery further upstream. 
However, due to the spatial distance, the CA bolus arrives later at the tissue of interest than suggested by the upstream measured AIF. Various studies have shown that neglecting this temporal delay decreases accuracy of the pharmacokinetic parameter estimates \citep{koh2011dce, Mehrtash2016bat, nadav2016bat}.
 
There are two standard approaches to determine the delay time: The first approach includes the time delay as an additional free parameter in the fitting process of the pharmacokinetic model \citep{meyer1989bat, kershaw2006bat}. The second type estimates the bolus arrival times (BATs) separately for the TCC and for the AIF; that is, the time points where the concentration curves start to rise. The time delay is the difference between the BAT of the TCC and that of the AIF. 
The advantage of this second type is that it does not require to include the time delay as an additional fit parameter which is likely to decrease fit stability.
A comparative study by \citet{kershaw2006bat} showed that the second type gives more accurate results than the first one. 

The gold standard for BAT estimation is the method of \citet{cheong2003automatic}
which is based on fitting a linear or convex quadratic function to the sample points in the neighborhood of the BAT.
An extension was proposed by \citet{singh2009bat} which gives comparable results.
These existing methods  were mainly developed for patient data which often exhibit fast upslopes (i.e. the peak of the signal is close to the BAT).
However, applying the method of
\citet{cheong2003automatic} to 
TCCs which do not exhibit this characteristic slope,
e.g. TCCs of small animals or weakly perfused tissues, may lead to poor results.
Another limitation is that estimated BATs have to coincide with one of the acquisition time points. This can lead to inaccurate estimates, in particular when working with low temporal resolution data.

Even though many DCE-MRI studies are performed in a preclinical environment employing small animals \citep{hectors2018dce, schreurs2017dce, glowa2017, keunen2011dce}, to our knowledge, there is no suitable approach for BAT estimation of DCE-MRI data of rats and mice. TCCs tend to vary dramatically depending on the underlying physiology and vascular architecture of the respective tissue ranging from steeply increasing and subsequently rapidly decreasing TCCs in e.g. intact vessels, to monotonously and shallowly increasing TCCs in necrotic areas. Hence, a more flexible model is required, able to adapt to various kinds of CA uptake curves. 

In this work we propose a novel method for BAT estimation of DCE-MRI TCCs that do not have a fast upslope -- as often observed in small animal data and patient data of tissues without a significant intravascular fraction. The method is based on approximation of concentration curves by splines which allow to adapt to a large variety of curve shapes after the BAT. The BAT itself and the involved secondary model parameters are
estimated automatically by generalized cross validation. 
Furthermore, the method is able to estimate BATs that do not coincide with one of the sampling time points.
An extensive study using simulated data and \textit{in vivo} rat data shows that the proposed method gives more accurate BAT estimations than the state-of-the art method.

\section{Materials and Methods}

\subsection{The proposed method}

\subsubsection{TCCs and BATs}

DCE-MRI measurements are  performed by acquiring a series of images at time points $t_n$ ($n=1,2, \ldots, N$). 
 After a number of images, the CA is given as a bolus. In this paper, it is assumed that the MR-signal intensities are already converted into concentrations of the CA.
Thus, the given data $c_n$ are
noisy samples of a
true TCC $C(t)$:
\[
    c_n
    = C(t_n) + \epsilon_n, \qquad\text{for } n = 1, \ldots, N.
\]
Here, the $\epsilon_n$ represent error terms
which are assumed to be independent and identically distributed Gaussian distributed random variables with zero mean and unknown variance.
For simplicity, we focus on the often used case of uniformly sampled time points, i.e., $t_n = (n-1) \Delta t,$ noting that the proposed method could also be extended to the more general case employing unevenly sampled time points.

Our goal is to estimate
the BAT $\tBAT$ 
from the noisy data $c_n$ 
which is the 
time point where the true TCC, $C(t),$ starts to rise.
Note that $\tBAT$ does not necessarily coincide with one of the sample time points $t_n.$

The BAT naturally divides a TCC into two parts. Prior to bolus arrival, the TCC is constant-valued at zero (unless previous boluses were injected),
whereas after the BAT, it rises and adopts different shapes depending on the underlying physiology. 
A common type of TCCs reaches their maximum almost immediately after the bolus arrival which is exploited in the method of \citet{cheong2003automatic}. But often the time between the BAT and the curve maximum is much longer, e.g. in small animal data or patient data of tissues without a significant intravascular fraction. Thus, building on the above assumption of a steep initial signal slope can lead to poor estimates of the BAT.
In the following,
we present a non-parametric approximation model based on smoothing splines that allows for adaption to various curve shapes,in particular for TCCs without fast upslopes. We note that the proposed method is also applicable to the raw MR data before conversion to concentration.

\subsubsection{Smoothing splines}

Smoothing splines are a classical signal estimation method
which have been applied to various biomedical problems \citep{craven1978smoothing,unser2002splines}.
We give a brief description here (for further details, the reader is referred to the references \citet{reinsch1967smoothing,deboor1978practical,wahba1990spline,unser1999splines}). 
First, recall that an interpolating spline of order $k$
is a function $u$ that passes through each data point
while minimizing its smoothness costs, 
as represented by the square integral of its $k$-th derivative
$\int_{t_1}^{t_N} (u^{(k)}(\tau))^2\, d\tau.$
It has been shown that 
an interpolating spline is a piecewise function
consisting of polynomials of maximum degree $2k-1$ on $[t_n, t_{n+1}],$ for $n = 1, \ldots, N-1$, 
such that all derivatives up to order $2k - 2$ coincide at the interface nodes $t_n.$
As we work with noisy data, 
we are interested in  approximation rather than in interpolation.
Instead of interpolating, 
a smoothing spline approximates the data by optimizing a tradeoff between smoothness
and fidelity to the data, as represented by the residual sum of squares.
More precisely, a smoothing spline of order $k \in \N$ is the solution of the following optimization problem:
Find a sufficiently smooth function $u$ on $[t_1, t_N]$
that minimizes the cost function
\begin{equation}\label{eq:splines}
   \sum_{n=1}^{N} (u(t_n) -c_n)^2
    + \alpha\int_{t_1}^{t_N} (u^{(k)}(\tau))^2\, d\tau,
\end{equation}
where $u^{(k)}$ denotes the $k$-th derivative of the function $u.$
The parameter $\alpha > 0$ adjusts the  relative weight of data fidelity and smoothness:
the larger the value of $\alpha$, the smoother the resulting spline.
To understand the role of the order $k,$ 
it is instructive to look at the solutions for very large $\alpha$:
For $\alpha \to \infty,$ the solution of \eqref{eq:splines} 
yields an ordinary polynomial of degree $k-1$ on $[t_1, t_N].$
This is because  the smoothness 
cost (the second term in \eqref{eq:splines}) 
is equal to zero exactly for polynomials of maximum degree $k-1.$
Thus, for $\alpha \to \infty,$ the minimizer of \eqref{eq:splines} coincides with the least squares approximation by a polynomial of degree $k-1.$

\subsubsection{Proposed approximation model}

\begin{figure}
    \centering
    \includegraphics[width=0.7\textwidth]{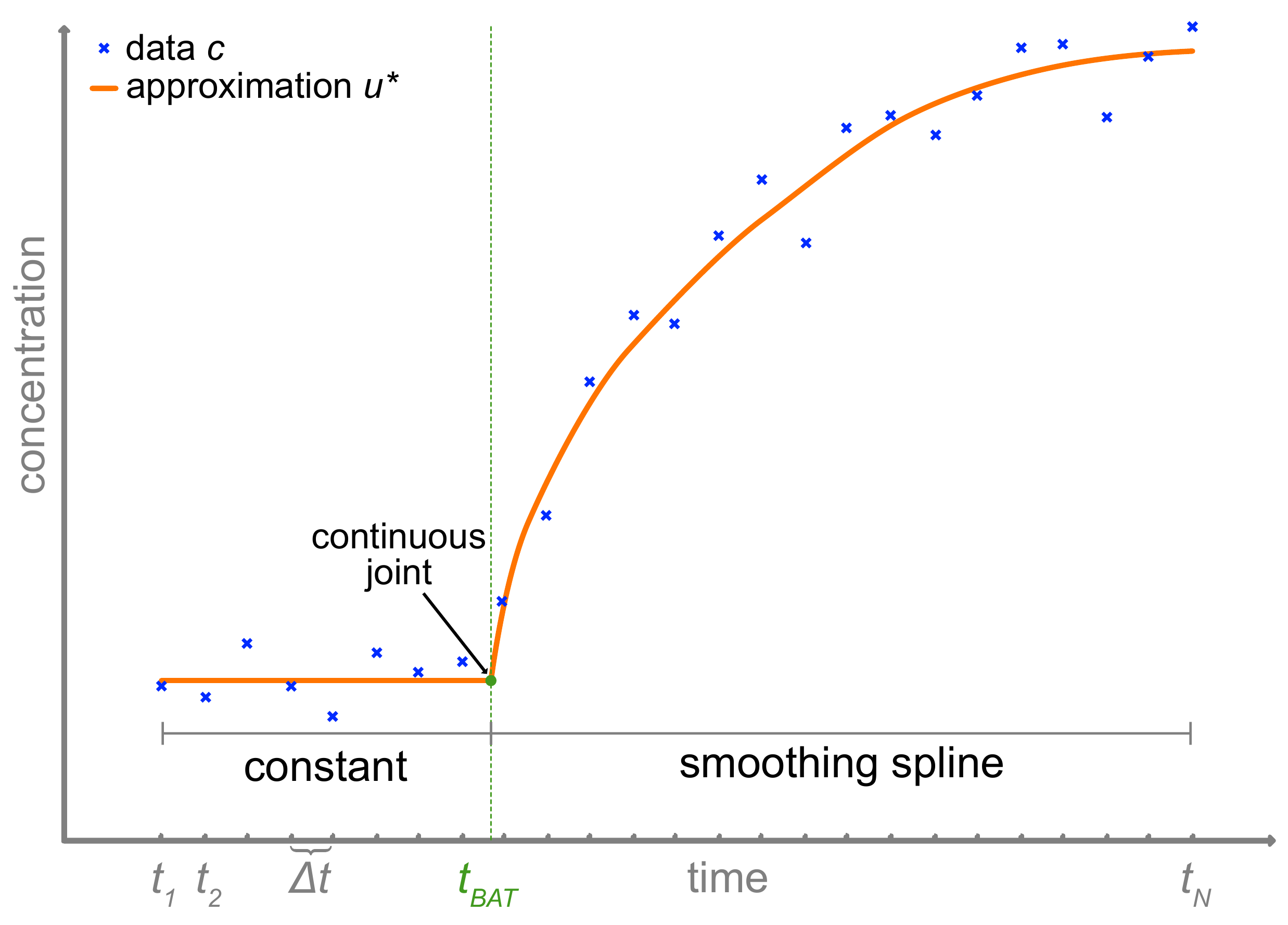}
    \caption{Graphical illustration of the proposed approximation model \eqref{eq:spline_model}. The given noisy TCC data (blue crosses) are approximated by a continuous piecewise defined function $u^*.$
    It is a constant function before the BAT $t_\text{BAT}$
    and afterwards a smoothing spline which is able to adapt to the shape of the curve by adjusting its stiffness parameters $\alpha$ and $k.$
    An estimate for the BAT, $t_\text{BAT},$
    is determined automatically 
    along with suitable spline stiffness parameters $\alpha$
    and $k$ by  generalized cross validation.
    Note that $t_\text{BAT}$ does not need to coincide with one of the sample times $t_1, \ldots, t_N.$} 
    \label{fig:illustration_model}
\end{figure}

The above discussion on the characteristics of a concentration curve motivates to approximate a TCC
by a continuous function that is constant on the interval $[t_1,\tBAT]$
and that is a smoothing spline on $[\tBAT,t_N].$
A graphical illustration is given 
in Figure~\ref{fig:illustration_model}.
Thus, the approximating function $u^*$
is defined by the minimizing condition
\begin{equation}\label{eq:spline_model}
\begin{split}
    u^* &= \argmin_{u}~\sum_{n=1}^{N} (u(t_n) -c_n)^2
    + \alpha\int_{\tBAT}^{t_N} (u^{(k)}(\tau))^2 \,d\tau. \\
    &\text{subject to $u$ is constant on $[t_1,\tBAT].$}
\end{split}
\end{equation}
As the approximation $u^*$ is not required to be smooth at the joint  it typically has a kink at $\tBAT$.

Since $u$ is constant on $[t_1, \tBAT],$ problem \eqref{eq:spline_model} can be reduced to
\begin{equation}\label{eq:spline_model_2}
\begin{split}
    \tilde u^* = \argmin_{\tilde u}~&\sum_{n=1}^{N'} (\tilde u(\tBAT) -c_n)^2 + 
    \sum_{n=N'+1}^{N} (\tilde u(t_n) -c_n)^2 
    + \alpha\int_{\tBAT}^{t_N} (\tilde u^{(k)}(\tau))^2 \,d\tau,
\end{split}
\end{equation}
where $N'$ is the index of the last time frame 
before $\tBAT$; that is, $N' = \max\{n: t_n \leq \tBAT\}.$
There is a simple relation between the
solutions of the constrained variant \eqref{eq:spline_model} and
that of the unconstrained variant \eqref{eq:spline_model_2}: $u^*(t) = \tilde u^*(\tBAT)$ 
for $t \in [t_1,\tBAT]$ and $u^*(t) = \tilde u^*(t)$ for  $t\in (\tBAT,t_N].$

Our principal interest is to estimate the parameter $\tBAT$ from the concentration curve $c.$ Alongside the estimation of $\tBAT$, the spline parameters $\alpha$ and $k$ have to be determined as well because these are needed for adapting the model to the curve shape after the BAT. All three parameters are estimated automatically as explained in detail in Section~\ref{sec:GCV}.

\subsubsection{Discretization and numerical computation of the approximating function}

To numerically compute the approximating functions, 
we discretize the optimization problem \eqref{eq:spline_model_2}.
First,  a finite difference approximation
of the $k$-th derivative $u^{(k)}$
is employed.
A natural choice for the set of discretization nodes $q$ are
the BAT and the subsequently sampled time points, i.e.,
$q = (\tBAT, t_{N' + 1},  t_{N' + 2}, \ldots, t_N).$
Let $\tilde v$ be the vector that contains the ordinates of $\tilde u$
 at these discretization nodes; that is, $\tilde v_i = \tilde u(q_i)$
 for $i = 1, \ldots, I$ with $I = N - N' + 1.$ 
A first order finite difference approximation 
of the $k$-th derivative is given by
\begin{equation}\label{eq:fd}
    \tilde u^{(k)}(q_i) \approx \sum_{j=0}^{k-1} \omega^{(k,\tBAT)}_{ij} \, \tilde v_{i+j}, \text{ for }i=1, \ldots, I - k,
\end{equation}
 where  the $\omega^{(k,\tBAT)}_{ij} \in \R$ are weights which we specify in the following.
First note that the weights  depend on $\tBAT$ because the first discretization node coincides with $\tBAT.$
Determining the weights $\omega^{(k,\tBAT)}_{ij}$ is a standard procedure; 
see e.g.~\citet{fornberg1996practical}. 
Further, we approximate the integral in \eqref{eq:spline_model_2} by a Riemannian sum.
In summary, we  obtain the discretized problem
\begin{equation}\label{eq:modelReduced}
    \begin{split}
  \tilde v^* = \argmin_{\tilde v \in \R^I}~&\sum_{n=1}^{N'} (\tilde v_1 -c_n)^2 + \sum_{n=N'+1}^N (\tilde v_{n-N'+1} -c_n)^2 \\ &+ \alpha\sum_{i=1}^{I-k} (q_{i+1} -q_i) \Big(\sum_{j=0}^{k-1} \omega^{(k,\tBAT)}_{ij} \tilde v_{i+j}\Big)^2.
  \end{split}
\end{equation}
Just as $\tilde v^*$ is a discrete approximation to
the continuous function $\tilde u^*,$
the expanded vector 
$v^* = (\tilde v^*_1, \ldots, \tilde v^*_1, \tilde v^*_2, \ldots, \tilde v^*_I )$ (of length $N$)
is a discrete approximation to $u^*.$

Equation \eqref{eq:modelReduced} is a least squares problem in $\tilde v.$
It is convenient to reformulate it in matrix notation,
\begin{equation}\label{eq:modelReducedLinear}
    \tilde v^* = \argmin_{\tilde v \in \R^I} \| B \tilde v -c \|^2 + \alpha \| W\Omega \tilde v \|^2,
\end{equation}
where $\|v\|^2$ denotes the squared Euclidean norm, $\|v\|^2 = \sum_{n} v_n^2,$
and the involved  matrices are as follows.
$W$ is a
diagonal matrix with the entries $W_{ii} = \sqrt{q_{i+1} - q_i}$ for $i = 1,\ldots,I-k.$ 
The matrix  $B \in \R^{N \times I}$ is defined by
\[
  B = \begin{bmatrix}
    \mathbf{1}_{N'-1}\, e_1^T  \\
    E_I 
  \end{bmatrix},
\]
where $\mathbf{1}_{N'-1} = (1, \ldots,1)^T$ denotes a vector of $(N'-1)$ ones, $e_1^T = (1, 0, \ldots, 0)$ 
is the first unit vector in $\R^{N'-1}$
and $E_{I}$ is the identity matrix of size $I.$
Further, $\Omega$  
implements the discrete approximation to the $k$-th derivative given in  \eqref{eq:fd}.
It is explicitly given by $\Omega_{i,i+j} = \omega^{(k,\tBAT)}_{ij}$ 
for $i = 1, \ldots, I,$ $j = 0, \ldots, k-1,$ and zero otherwise.
For readability, we omit the dependence of the involved matrices
on $\tBAT$ and $k$ in our notation.

A standard computation (differentiation of the functional in \eqref{eq:modelReducedLinear} with respect to $\tilde v$ and setting it equal to zero) reveals that $\tilde v^*$ is the solution of the linear system
\[
    ( B^T B  +  \alpha (W \Omega)^T W \Omega) \, \tilde v^* = B^Tc.
\]
Solving this equation for $\tilde v^*$
and using the relation $v^* = B \tilde v^*$
gives the explicit solution
\begin{equation}\label{eq:linear_estimator}
    v^* = H^{(\tBAT, \alpha, k)}c
\end{equation}
where  $H^{(\tBAT, \alpha, k)} \in \R^{N\times N}$ is the influence matrix given by
\begin{equation}\label{eq:linfluence_matrix}
    H^{(\tBAT, \alpha, k)} = B
    (B^T B  +  \alpha (W \Omega)^T W \Omega)^{-1}  B^T.
\end{equation}
Note that $H$ depends on $\tBAT$, on $\alpha$, and on $k$ because the right hand side in \eqref{eq:linfluence_matrix} does so.
Thus, for fixed values of $\tBAT,$ $\alpha,$ and $k,$
the discrete approximation ${v}^*$ of $u^*$ 
is given by multiplication of the data vector $c$
with the matrix $H^{(\tBAT, \alpha, k)}.$

\subsubsection{Parameter estimation by generalized cross validation}\label{sec:GCV}

Our principal interest is to determine the BAT by estimating the parameter $\tBAT.$
For this, 
the spline parameters $\alpha$ and $k$ 
need to be estimated as well because
these affect the adaption of the model to the curve shape after the BAT.
A widely used approach for determining such parameters
is generalized cross validation (GCV) \citep{craven1978smoothing, golub1979generalized},
which has shown to provide good performance for spline-based estimators 
\citep{wahba1985comparison, aydin2006empirical}.
The GCV score 
is a ratio of the mean squared error (MSE), given by $\| v^* -c \|_2^2/N,$
and a quantity based on the degrees of freedom ($\df$) of the estimator.
As the proposed model leads to a linear estimator,
the $\df$ coincide with the trace of the influence matrix $H^{(\tBAT, \alpha, k)}$.
Thus the generalized cross validation score for the proposed model has the explicit form
\begin{equation}\label{eq:GCV}
    \GCV(c; \tBAT, \alpha, k) = \frac{\operatorname{MSE}}{(1 - \frac{1}{N}\df)^2} =  
    \frac{\frac{1}{N}  \| H^{(\tBAT, \alpha, k)}c - c \|_2^2}{ (1 - \frac{1}{N}\trace H^{(\tBAT, \alpha, k)})^{2}}.
\end{equation}
The optimal set of parameters $(\tBAT^*, \alpha^*, k^*)$ in the sense of
GCV is the one that provides the minimum GCV score:
\begin{equation}\label{eq:minGCV}
(\tBAT^*, \alpha^*, k^*) = \argmin_{\tBAT, \alpha, k} \GCV(c; \tBAT, \alpha, k),
\end{equation}
where $\tBAT$ is searched in the continuous interval $[t_1,t_{N-k}],$ $\alpha$ in the continuous interval $[0, \infty),$ and $k$ in a set of positive integers $\{k_{\min{}}, \ldots, k_{\max{}}\}.$
A minimizer $\tBAT^*$ is the proposed estimate of 
the BAT. 

It is possible that \protect\eqref{eq:GCV} has multiple minima meaning that there are multiple equally good candidates for $\tBAT^*$ in the sense of the proposed model. In such a scenario -- which is the exception rather than the rule --  the utilized minimization algorithm implicitly takes the decision for one of the minima.

\subsubsection{Computational minimization of the GCV score and implementation}

All computational work was performed in MATLAB R2017b (MathWorks, Inc. Natick, Massachusetts, USA).
The finite difference weights 
in \eqref{eq:fd} are computed using the implementation \enquote{fdweights} of T.~Driscoll available at the MATLAB File Exchange.\footnote{Available at \url{ https://de.mathworks.com/matlabcentral/fileexchange/13878-finite-difference-weights}, last checked on Nov. 12, 2018.}

Regarding the minimization of \eqref{eq:minGCV} with respect to the spline order $k,$ we first discuss the role of that parameter:
 By construction, a spline of order $k$ 
 receives no smoothness penalty when approximating data by a polynomial of degree $k-1.$
 This means that long-term polynomial trends of order $k-1$ 
 are particularly well captured. (However, note that splines are more flexible than polynomials.)
 The long-term trends of the considered types of TCCs 
 are beyond linear which suggests to choose
 $k_{\min{}} > 2.$
 On the other hand, approximation with high-order polynomials 
  is prone to overfitting, especially in the case of high noise levels, and may get ill-conditioned, in particular for longer signals.
 For the considered types and lengths of TCCs, we found that $k_{\max{}} \leq 6$ leads to stable solutions.
 To improve the stability in case of even higher noise  or even longer TCCs, one can consider reducing  $k_{\max}$ to $5$ or $4.$
 Summing this up, we found $ k_{\min} = 3 $ to $k_{\max} = 6$ to be a reasonable range for the signals considered in this work.
 
To minimize the GCV score with respect to the two continuously defined variables $\tBAT$ and $\alpha,$
 MATLAB's solver \enquote{fminunc},
 an iterative solver
 based on Newton's method, is used with the logarithm of the GCV score as target function. (This is equivalent to minimizing the original GCV score as the logarithm is strictly monotone.) We use the variant based on numeric differentiation
 since deriving analytic expressions for the Jacobian and the Hessian of the GCV score is involved. 
To avoid getting trapped in a local minimum, we first perform a coarse search over a grid of $\alpha$ and $\tBAT$
and use the optimum values as starting point for the iterative solver.
To save computing time, we limit the coarse search interval for $\tBAT$ to the interval 
between the time points $t_{\min{}}$ and $t_{\max{}}$ which correspond to the minimum and the maximum of the signal. 
Although this restriction did not negatively affect the results for the TCCs considered in this work,
one could consider dropping this restriction
when processing data with even stronger noise
 so that the coarse search space does not get restricted in a disadvantageous way. Note that the fine search always uses
 the full search space $[t_1, t_{N - k}].$
 If the injection time of the CA is known,
it can be used to obtain an improved lower boundary for $\tBAT.$
For the evaluation in this work, we do not assume that the  injection time is known,
so that the method is applicable even if the injection time has not been recorded.
To prevent the iterations to run out of the search domain
 for $(\tBAT,\alpha)$  barrier functions are added.
 These are functions 
 that are zero in the interior of the search domain
 and grow fast and smoothly to infinity when approaching the boundaries.
 Further, computing the influence matrix $H$
 based on the normal equation in \eqref{eq:linfluence_matrix}
 can be numerically instable when working with higher spline orders and with longer signals.
 To avoid this we perform the computation based on
 the least squares formulation in \eqref{eq:modelReducedLinear} 
 using the QR solver implemented in MATLAB.
  We also observed that very small values
 for $\alpha$ often led to overfitting.
 To prevent this, we limit the  search
 range for  $\alpha$ by a minimum value $\alpha_{\min}$ which was set to $\alpha_{\min} = 1$ throughout this work. A pseudocode of the proposed method is given in Appendix A. 
 Our reference implementation 
 is provided online.\footnote{Reference implementation available at \url{https://github.com/mstorath/DCEBE}.}

\subsection{Validation using simulated data}

For validation, the proposed method was applied to simulated TCCs and the respective AIFs with known true BAT. 

\subsubsection{Data simulation}\label{sec:data_sim}

Data were simulated using an in-house developed tool within the \enquote{Medical Imaging Interaction Toolkit (MITK)} \citep{Nolden2013, debus2017dce}. TCCs  were generated by forward convolution of an AIF with the respective tissue response function of a pharmacokinetic compartment model.
We used the tissue response functions of two established pharmacokinetic models. The two compartment exchange model (2CXM) \citep{sourbron2012pkm} is characterized by four parameters, the plasma flow $F_\text{p}$, the permeability $PS$, and the fractional volumes of the extravascular extracellular compartment  $v_\text{e}$ and the plasma compartment $v_\text{p}$. In the Extended Tofts Model (ETM) \citep{tofts1999etm, sourbron2011etm} the parameters $F_\text{p}$ and $PS$ are combined to the transfer constant $K_{\text{trans}}$. 

TCCs were simulated for small animals, more precisely, rats, using the model-based AIF proposed by \citet{mcgrath2009rataif} (model B). Three physiologically realistic TCC types were simulated for both pharmacokinetic models by varying the input parameters (ETM simulated: $K_{\text{trans}}$ = 15 ml/min/100ml, $v_\text{p}$ = 0.05, and $v_\text{e}$ = 0.3, 0.6, or 0.1, respectively; 2CXM simulated: $F_\text{p}$ = 25 ml/min/100ml, $PS$ = 10 ml/min/100ml, $v_\text{p}$ = 0.05, and $v_\text{e}$ = 0.05, 0.15, or 0.30, respectively).

To show the flexibility of the proposed method, patient data was additionally simulated, using a population-average based AIF proposed by \citet{parker2006aif}. Here, two TCC types were investigated (ETM simulated: $K_{\text{trans}}$ = 10 ml/min/100ml, $v_\text{p}$ = 0.1, and $v_\text{e}$ = 0.05 or 0.3, respectively; 2CXM simulated: $F_\text{p}$ = 25 ml/min/100ml, $PS$ = 5 ml/min/100ml, $v_\text{p}$ = 0.1, and $v_\text{e}$ = 0.15, or 0.05, respectively).

\begin{figure}[t]
    \centering
    \includegraphics[width=1\textwidth, trim=0 0 100 0,clip]{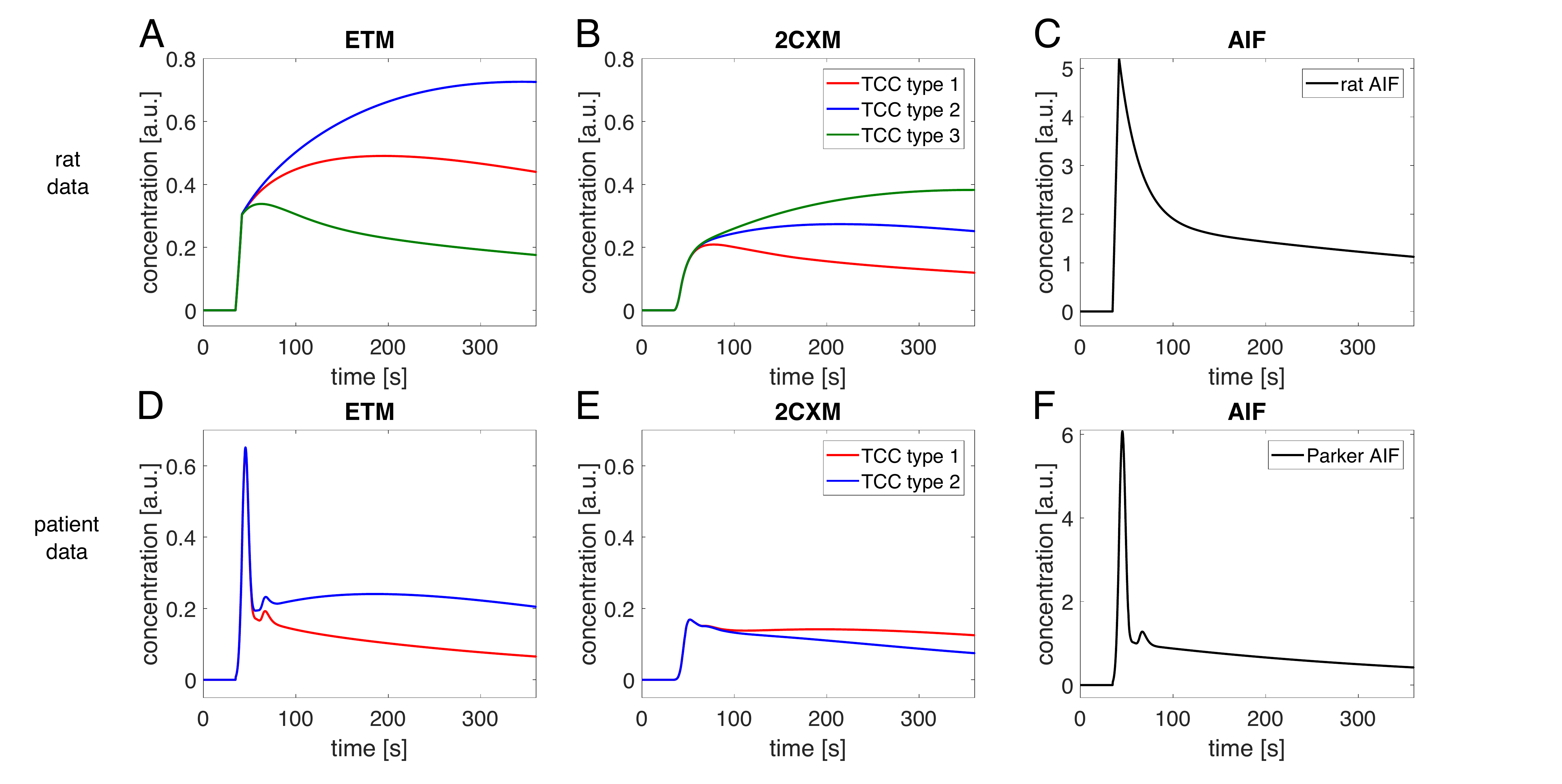}
    \caption{Simulated concentration curves used for validation of the proposed method: (A)-(B) Rat TCCs simulated with ETM and 2CXM. (C) Model-based rat AIF as proposed by \citet{mcgrath2009rataif}. (D)-(E) Patient TCCs simulated with ETM and 2CXM. (F) Population-average based Parker AIF \citep{parker2006aif}. All concentration curves are displayed with their original temporal resolution of 0.25 s.}
    \label{fig:original}
\end{figure}

High resolution TCCs were simulated for an acquisition time of 360 s with a temporal resolution of 0.25 s and BAT at 34.75 s (Figure \ref{fig:original}). BAT was defined as the last point on the concentration curve’s baseline.  

\subsubsection{Estimation of BAT on simulated data for different temporal resolutions and SNRs}

To evaluate the robustness of the proposed method, simulated TCCs were altered for four temporal resolutions ($\Delta$t = 1 s, 2 s, 5 s, 7 s) and four noise levels defined by their signal-to-noise ratio (SNR) (SNR = 100, 50, 25, 10). SNR was defined as the ratio between the maximum of the concentration curve to the standard deviation of the Gaussian noise added to the concentration curve. Different temporal resolutions were realized by respectively down-sampling the original data. BATs were estimated by the proposed method for 1000 noise realizations for each configuration (curve type, SNR, temporal resolution) to obtain statistical information.

Results were compared to those obtained by the method of \citet{cheong2003automatic} who are using a linear-quadratic piecewise continuous function for approximating the concentration-time curve and estimating the BAT. (A more detailed description of Cheong's method and its mathematical relation to the proposed method is given in Appendix B and C). To our knowledge this method is still the current gold standard for BAT estimation.

\subsection{Estimation of BAT on \textit{in vivo} rat data}\label{sec:mm_est_invivo}

To validate our approach for \textit {in vivo} data, DCE-MRI data of five tumor-bearing Copenhagen rats (Charles River Laboratories Inc., MA, USA) were investigated. All experiments were approved by the governmental review committee on animal care (Ref. No. 35-9185.81/G-88/14), and animals were kept under standard laboratory conditions. Briefly, animals were transplanted with fresh fragments of the syngeneic Dunning prostate adenocarcinoma subline R3327-HI \citep{isaacs1978tumor} in both distal thighs. When tumors reached a size of~10~$\times$~10~mm$^2$ they were imaged at a 1.5~T~MRI system (Symphony, Siemens, Erlangen, Germany) using an in-house built animal coil. Animals were anaesthetized by inhaling a mixture of 2~\% isoflurane (Baxter, Unterschließheim, Germany) and 2 l/min oxygen. T2-weighted images were acquired with a turbo spin echo sequence (TR~3240~ms, TE~82~ms, slice thickness: 1.5~mm, pixel size: 0.35~mm~$\times$~0.35~mm) for anatomy references. Two DCE slices (slice thickness: 4.5~mm, pixel size: 0.99~mm~$\times$~0.99~mm) were positioned transversely in the rat: one slice in the heart for acquisition of the AIF, the other one through the largest diameter of the tumors. DCE-images were acquired with a T1-weighted TURBO-FLASH sequence (TR 272 ms, TE 1.67 ms, flip angle 20$^\circ$) for a total acquisition time of 380 s with a temporal resolution of 0.75 s. 0.1 mmol/kg Gd-DTPA (Magnevist, Bayer Healthcare Pharmaceuticals Berlin, Germany) was manually injected \textit{i.v.} approximately 30 s after starting the DCE sequence. Even though the experimenters followed a strict routine for CA injection, some temporal variation for the time point of injection might have occurred between animals, as our experimental set-up only allowed to record the time point of entering the MR room and not the actual time point of injection.

DCE-data was converted to concentration curves in MITK by means of absolute signal enhancement ($C(t) = \kappa \cdot (S(t) - S(0))$, where $S$ is the MR signal at time $t$ and $\kappa$ delineates a proportionality which was set to 1).
The AIF was extracted from one voxel in the left ventricle of the rats' hearts with the steepest rise, highest signal amplitude, and least noise. The tumors were segmented based on the acquired T2-weighted images. TCCs were extracted voxel-wise for each tumor. BATs were estimated using the proposed method as well as the method developed by \citet{cheong2003automatic}. Additionally, we compared our results to an adapted version of Cheong's method, which only considers sample points after the CA injection time for BAT estimation (\emph{Cheong \textit{et al.} adapted)}. Because our experimental set-up did not allow for exact recording of the injection time, we set a plausible common lower boundary. As we can ensure that no CA was injected before 25 s, 
earlier time points were not considered for BAT estimation. 
Thus the adapted version of Cheong's method
 considered the sample point at 25.37~as the first one possible.

\section{Results}

\subsection{BAT estimation on simulated rat data}
\begin{figure}[t]
    \centering
    \includegraphics[width=1\textwidth, trim=160 70 200 140, clip]{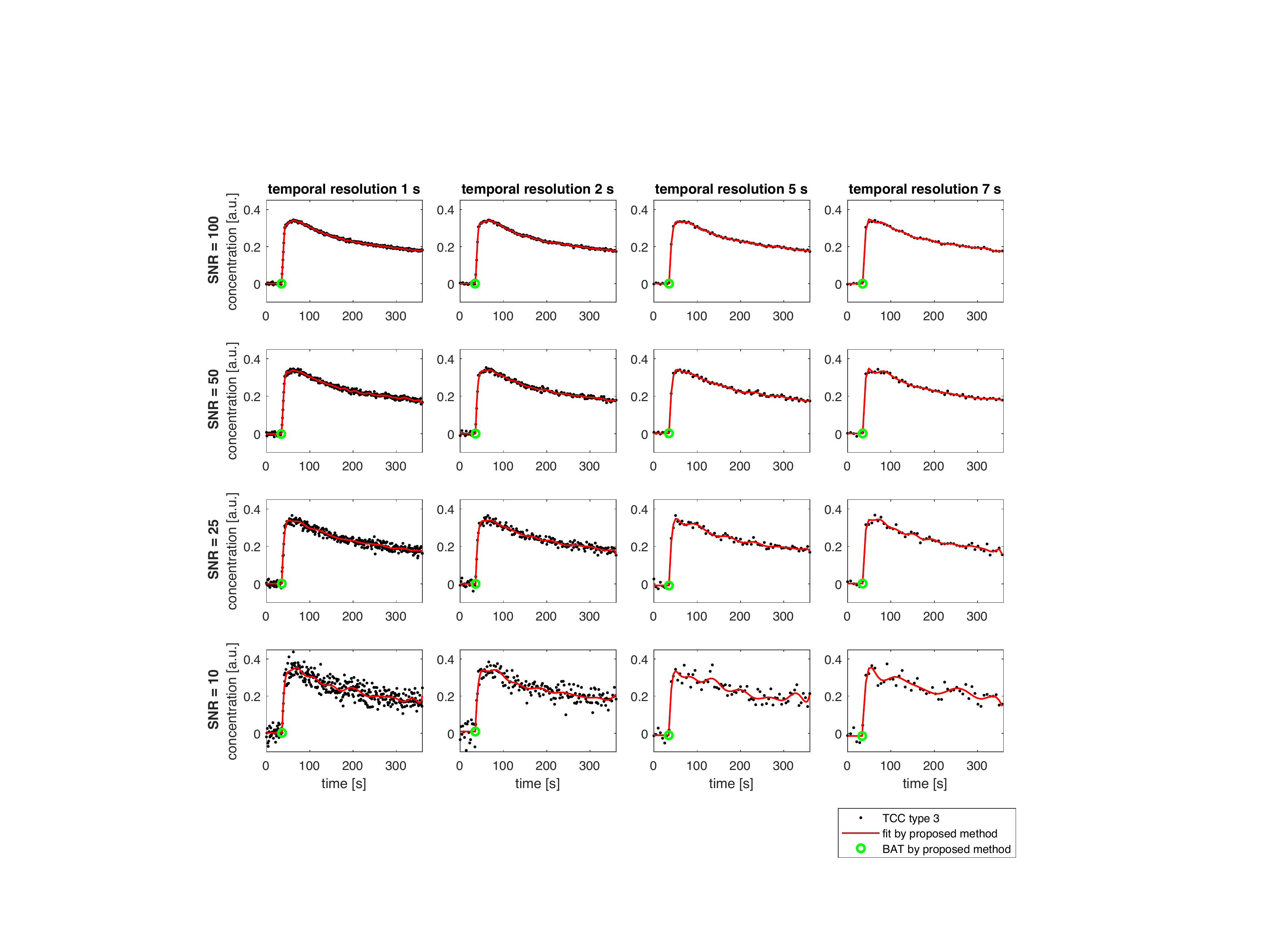}
    \caption{Exemplary results for BAT estimation in rats: Type 3 TCCs (black dots) simulated with the ETM are displayed for all temporal resolutions (columns) and SNRs (rows) together with the respective fit by the proposed method (red line) and estimated BAT (green circle).}
    \label{fig:resultsCT3}
\end{figure}

 \begin{figure}[t]
    \centering
    \includegraphics[width=1\textwidth,trim= 20 0 90 0, clip]{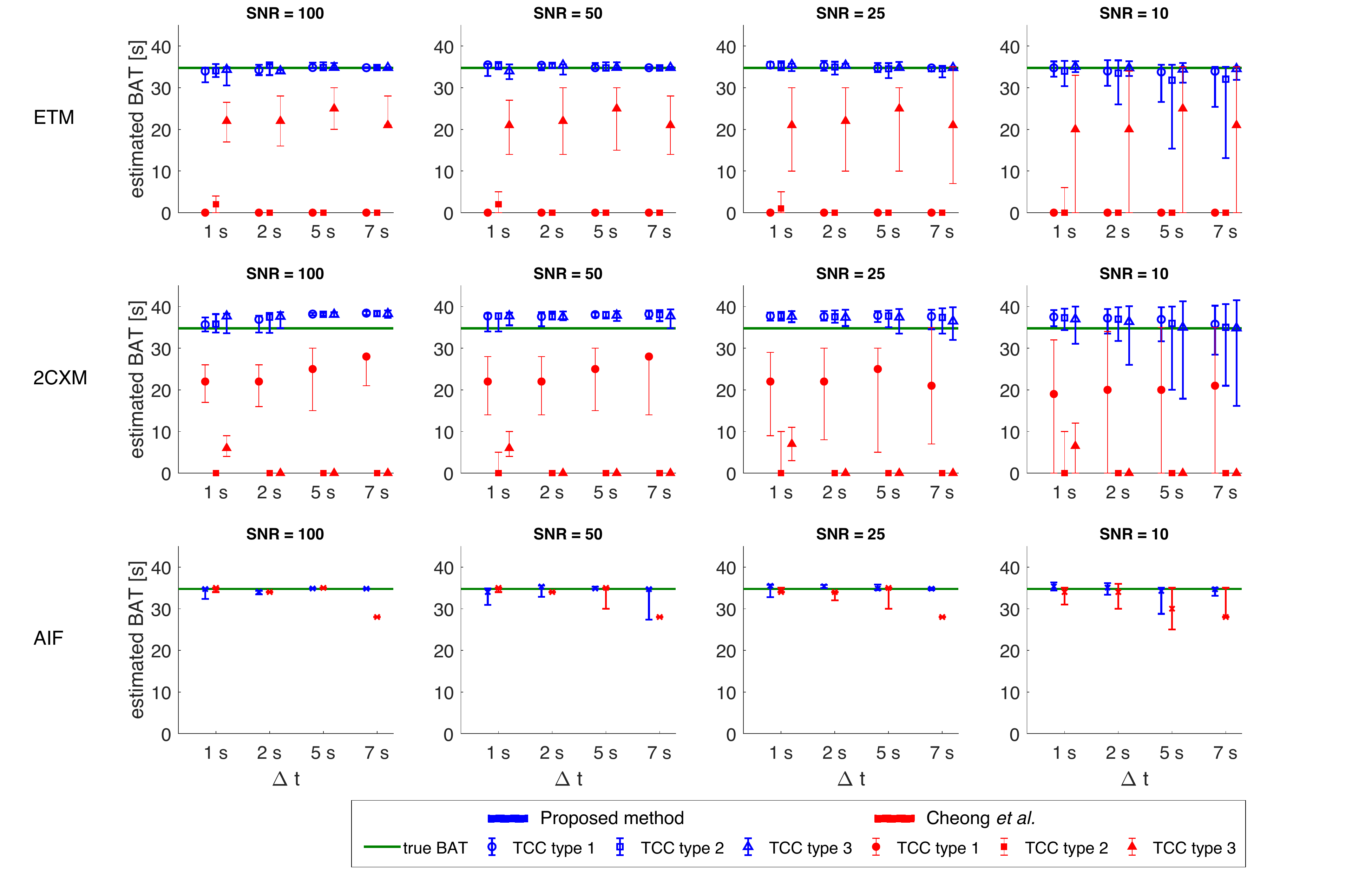}
    \caption{Illustration of the accuracy of BAT estimation for rat data: three TCC types were generated with the ETM and the 2CXM. Four temporal resolutions ($\Delta$t = 1 s, 2 s, 5 s, 7 s) and four noise levels (SNR  = 100, 50, 25, 10) were used on the simulated TCCs and AIFs. Results are presented as medians with 5 \% and 95 \% percentiles of 1000 repetitions of BAT estimation per TCC type with the proposed method (blue symbols) and Cheong's method (red symbols). The green line identifies the true BAT.}
    \label{fig:ratBat}
\end{figure}

Exemplary fits by the proposed method to one TCC type for all temporal resolutions and noise levels are displayed in Figure~\ref{fig:resultsCT3}. Figure \ref{fig:ratBat} displays the results of BAT estimation for various TCC types and AIFs for various temporal resolutions and SNRs representing different data quality. 

Cheong's method provides accurately estimated BATs for AIFs with high temporal resolutions. The accuracy is diminished for AIFs with low temporal resolutions.
For curves with a less significant peak in the vicinity of the BAT, such as TCC type 3 for the ETM and TCC type 1 for the 2CXM, the estimated BATs are relatively far off the ground truth and are scattered over a wide range. Cheong's method gives poor results for estimated BATs of the remaining TCCs. 

The proposed method gives comparable results for the AIFs at high temporal resolutions and performs better for the AIFs at the lowest temporal resolution. 
For the TCCs, the proposed method could precisely estimate BATs of all shapes and in most cases did not exhibit much variability as indicated by the small range between the  5 \% and 95 \% percentiles. A slight overestimation of BATs was observed for TCCs simulated with the 2CXM. This could be explained by the shallow initial rise in concentration observed for TCCs simulated with this model. Furthermore, the results are robust towards increased noise levels and low temporal resolutions of the input data.  
In almost all configurations the proposed method significantly improves the accuracy of BAT estimation upon the state-of-the-art method for the tested TCCs.

\subsection{BAT estimation on simulated patient data}

 \begin{figure}[t]
    \centering
    \includegraphics[width=1\textwidth,trim= 20 0 90 0, clip]{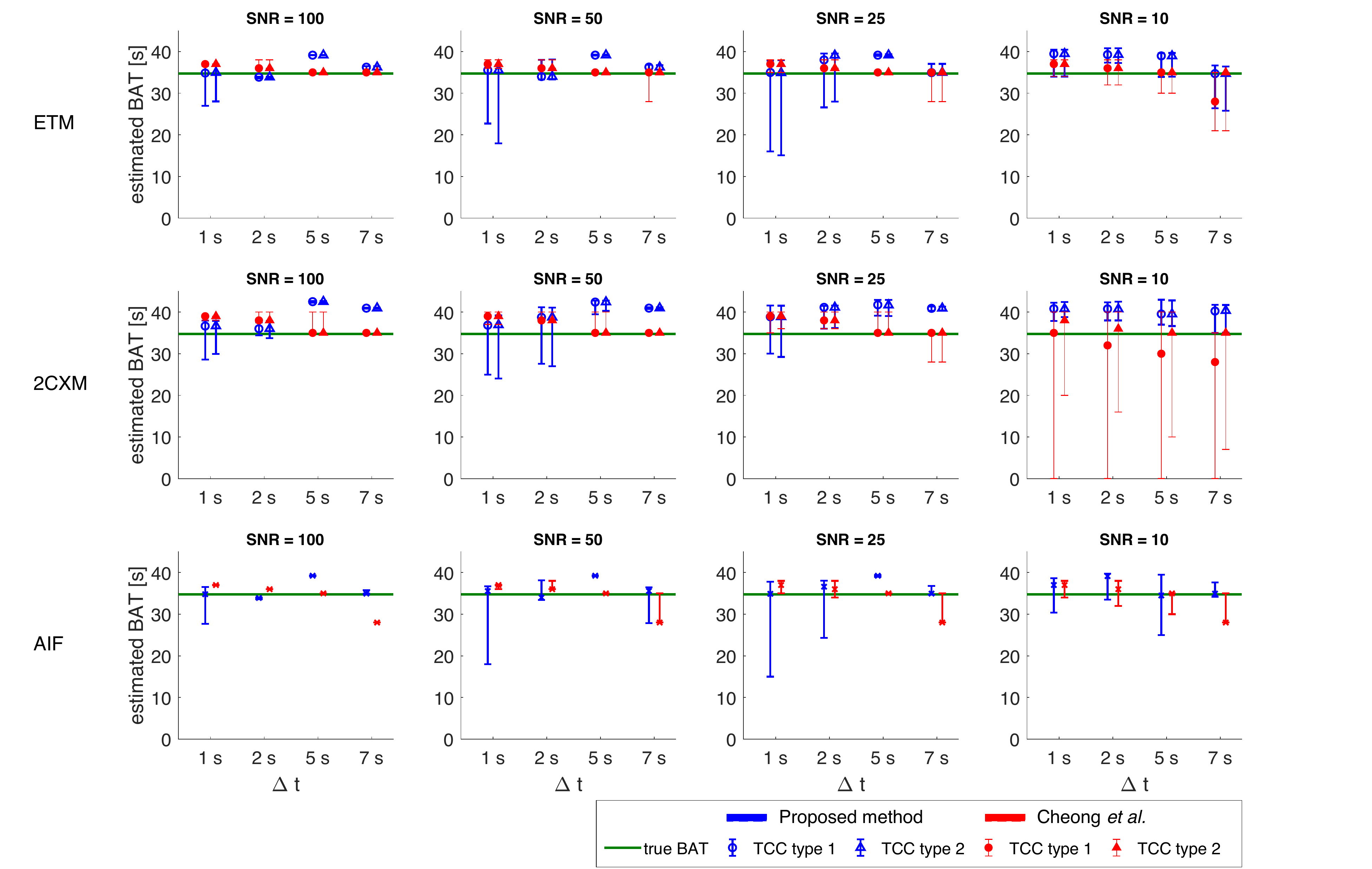}
    \caption{Illustration of the accuracy of BAT estimation for patient data: two TCC types were generated with the ETM and the 2CXM. Four temporal resolutions ($\Delta$t = 1 s, 2 s, 5 s, 7~s) and four noise levels (SNR  = 100, 50, 25, 10) were used on the simulated TCCs and AIFs. Results are presented as medians with 5 \% and 95 \% percentiles of 1000 repetitions of BAT estimation per TCC type with the proposed method (blue symbols) and Cheong's method (red symbols). The green line identifies the true BAT.
    }
    \label{fig:parkerBat}
\end{figure}

Human AIFs are characterized by a higher and steeper initial peak which might lead to different TCC shapes as compared to those of small animals (Figure \ref{fig:original}). Results of BAT estimation by both methods are displayed in Figure \ref{fig:parkerBat}.

Cheong's method gives accurately estimated BATs with only small variation for low temporal resolution TCCs. Some major variation in the obtained results can only be observed for data simulated with the 2CXM at the lowest SNR. Cheong's method performed equally well for all simulated patient TCCs.

The proposed method gives results close to the ground truth with a slight overestimation of BATs for low SNR TCCs and data simulated with the 2CXM.

Thus, for patient data, the proposed method gives competitive results, yet Cheong's method performs slightly better.

\subsection{BAT estimation on in vivo rat DCE-MRI data}

The proposed method was also tested on \textit{in vivo} data that was acquired from 10 tumors in 5 rats. BATs were estimated for each tumor voxel and the results are displayed in Table \ref{tab:invivo}. 
\begin{table}[]
\caption{
    Estimated BATs for \textit{in vivo} rat data (in seconds) by Cheong's orignal and adapted approach and the proposed method. For the tumor TCCs, the medians and 5 \% and 95 \% percentiles are reported. The second and the third column indicate the number of evaluated voxels (i.e. TCCs) in the right and in the left tumor, respectively. For Cheong's adapted method a plausible lower boundary of 25.37 s was chosen.}
    \label{tab:invivo}
    \vspace{1ex}
    \centering
   
    {\scriptsize
    \input{Table_1.txt}
    }
\end{table} 

\begin{figure}
    \centering
    \includegraphics[width=1\textwidth, trim=110 10 110 0, clip ]{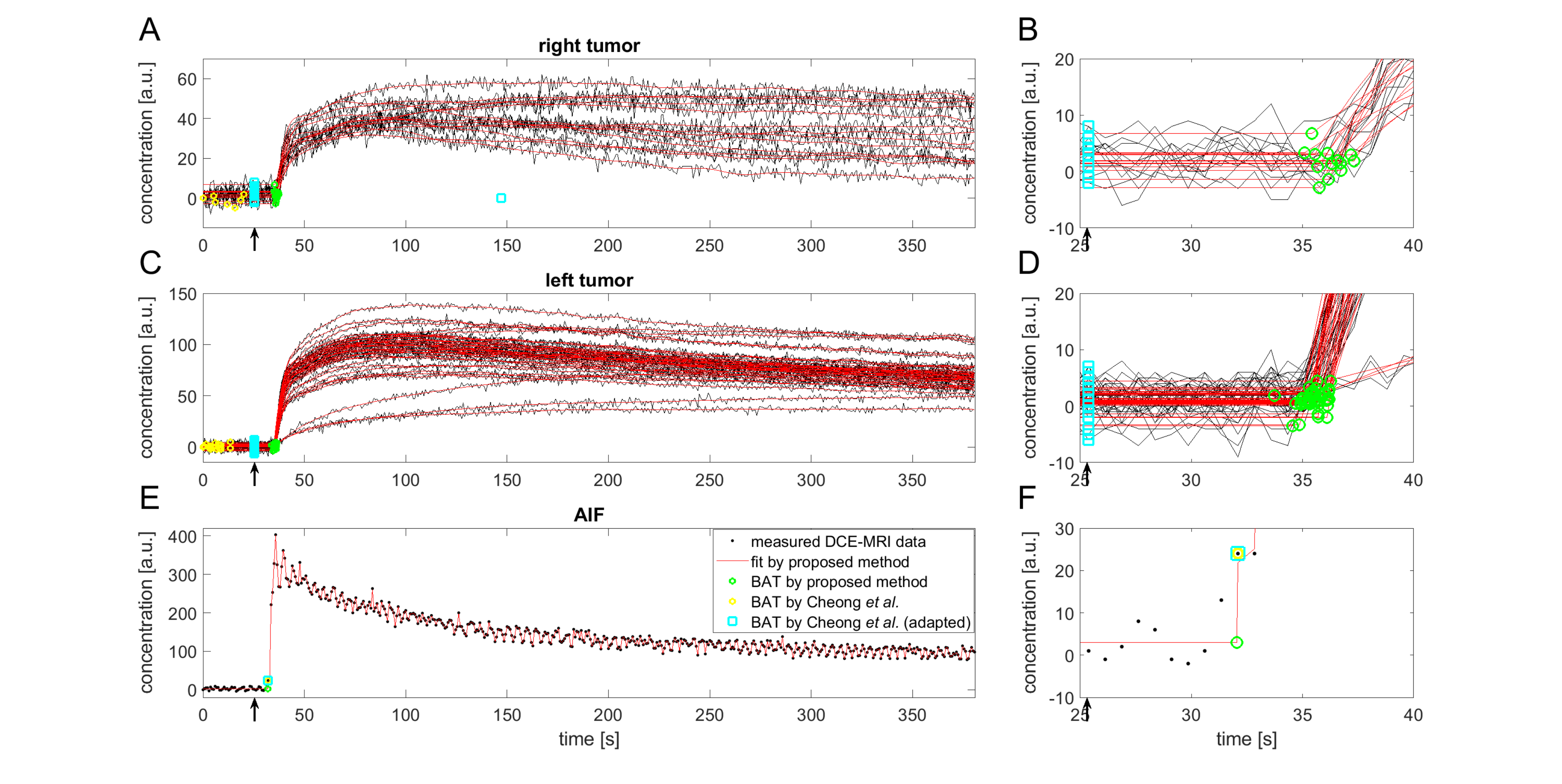}
    \vspace{-5ex}
    \caption{Exemplary pixel-based concentration curves of animal 4: (A) right tumor with close up of the BAT region (B); (C) left tumor with close up of the BAT region (D); (E) AIF and close up of the BAT region (F). BATs estimated by the proposed method, by Cheong's original approach and by Cheong's adapted approach are marked by green and yellow circles and blue squares, respectively. The black arrows indicate the lower boundary used for the injection times (needed for adapted version of Cheong's method).
    }
    \label{fig:inVivo}
\end{figure}

The compared methods estimated identical BATs for the rats' AIFs (Table~\ref{tab:invivo}, Figure~\ref{fig:inVivo}~E-F). However, for the TCCs the BATs estimated by Cheong's method are implausible as almost all of them were estimated to occur even before the lower bound of injection times. The adapted version of Cheong's method
which excludes implausible sample points below the lower bound of $25.37$~s for BAT estimation (see Section \ref{sec:mm_est_invivo})
resulted in this lower bound in most cases.
However, visual inspection reveals that the TCCs do not rise significantly before approximately 35 s.
This means that the results of the adapted version remain uninformative in most cases.

The BATs estimated by the proposed method are very close to each other within one animal and appear to be at reasonable time points by visual inspection (Table~\ref{tab:invivo}, Figure \ref{fig:inVivo} A-D). DCE-curves of animal 4 with estimated BATs for both tumors and AIF are displayed as exemplary results in Figure~\ref{fig:inVivo}. Despite the noticeable variation in TCC shapes occuring even within one tumor, the proposed method fits all curves equally well and estimates reasonable BATs. We emphasize that the proposed method 
does not require a lower bound of injection times as a priori information.

\subsection{Computation time}

On a laptop computer with 
2.2 GHz Intel Core i7, 16 GB RAM, MATLAB R2017, the computation time for the simulated TCCs with $N = 361$ time samples was as follows:
The computation time of the baseline approach \citep{cheong2003automatic}
is around $0.15$~seconds per signal curve. The proposed method took around $108$~seconds 
for processing a single TCC.
Processing $100$ TCCs simultaneously took around $347$~seconds,
i.e. around $3.5$~seconds per TCC in average.
The reason for the more favourable runtimes in simultaneous processing
is that a series of expensive matrix operations do not have to be recomputed for every signal.
Finally, we found that the optimal order $k^*$ of the smoothing splines resulted to be 5 or 6 in most cases.

\section{Discussion}

Estimating the time delay between the arrival 
of the CA in the upstream artery and the tissue of interest accurately
is an important step 
to obtain correct parameters from pharmacokinetic modelling of DCE-MRI data \citep{Mehrtash2016bat}. 
One way to reliably obtain the delay is to take the 
time difference between the estimates of the BATs of the TCC and of the AIF, respectively \citep{kershaw2006bat}.
The gold standard method for estimating BATs has been developed for TCCs with a fast upslope to the peak near the BAT as often observed in patient data.  However, we found that this method does not give accurate BAT estimations in small animal data and shallowly rising concentration curves and 
to our knowledge, there is currently no commonly accepted method addressing this problem. 
To close this gap, we have developed a novel method for BAT estimation in such data.
The proposed method  approximates the TCC 
by a constant function before the BAT and by a smoothing spline after the BAT.
Determination of the BAT and  adaption of the spline to the curve shape is done automatically by generalized cross validation.

We used simulated data with known ground truth to evaluate the accuracy of our approach and compared our results to the method by \citet{cheong2003automatic}. 
The results showed that Cheong's method works very well on patient data with fast upslopes but
gives unsatisfactory results for rat TCCs. In contrast,
the proposed method estimates BATs
that are close to the ground truth.
The main reason is that the method of \citet{cheong2003automatic}
heavily relies on a fast upslope from the BAT to the
peak of the concentration curve which is frequently not the case in small animal data.
A central advantage of the proposed method is
that it does not need such a special curve characteristic.
Instead, the smoothing spline is able to adapt to a variety of 
curve shapes after the BAT.

The arrival of the CA does typically not coincide with one of the  time points of the image acquisition.
The state-of-the-art method, however, 
limits estimation of BATs to those time points
which imposes a systematic upper limit to the estimation accuracy.
The proposed method estimates BATs on a continuous scale;
i.e.~the estimated BAT can lie in between the time points of the image acquisition. 
This improves the accuracy of the results especially for data with low temporal resolution. 

Even though the proposed method was developed for small animal data, 
we found that it is flexible enough to be used for BAT estimation in patient data as well:
For the higher temporal resolutions and higher SNRs,
the experimental results are competitive 
with those obtained by the method of \citet{cheong2003automatic}.
Yet, for lower temporal resolutions and lower 
SNRs Cheong's method gave more accurate estimates of the BATs in patient data.

The proposed method was also applied to real \textit{in vivo} acquired rat data. As the ground truth is not known for this data we can only make qualitative statements on the results. Cheong's approach gave reasonable results for the AIFs which fulfill the criterion of a fast upslope from  the BAT to the concentration maximum.
For the TCCs, however, the estimated BATs  can be considered as implausible because they laid before the actual injection time.
Adapting Cheong's method to consider only BATs 
after a plausible lower time point mostly resulted in that lower bound.
In either case, the estimates for the BATs are mostly not informative. In contrast, the BATs estimated for the TCCs by the proposed method
were found to be very close to each other, for each tumor and each animal, respectively. When evaluating the location of the BATs within the TCCs, they appear to be set at reasonable positions.

Due to the complexity of the model, the computational effort of the proposed method is higher than that of the method of Cheong, especially for long signals. 
Generally, the average computation time per signal
can be dramatically reduced when processing multiple signals simultaneously (for example, all voxels of all the imaged objects).
Computational time can be reduced further
at a potential trade-off in accuracy:
First the search space for the order $k$ can be restricted to the values 5 and 6 as we observed them to be the optimal order $k^*$ of the smoothing splines in most cases. Second, the signal tails could be cropped to reduce the length of the signal.

\section{Conclusion}

We proposed a novel method for BAT estimation in DCE-MRI data
which is able to adapt to various types of TCCs; in particular it does not require the TCCs to have a fast upslope from the BAT to the peak.
The method is based on a flexible spline-based approximation model and automatic parameter estimation by generalized cross validation. We evaluated the method on simulated data (rat and patient) with known ground truth as well as on \textit{in vivo} rat data. The proposed method was found to be more accurate in rats than the method of \citet{cheong2003automatic}, which has been developed for the analysis of patient data with fast upslopes. Furthermore, the proposed method even gave competitive results in the considered patient data.
We propose using this method for reliable BAT estimation of DCE-MRI data that does not have a fast upslope -- as often observed in small animal data and patient data of tissues without a significant intravascular fraction -- to correct the delay between CA arrival in the AIF and the tissue of interest.

\section*{Acknowledgement}

This work was supported by the German Research Foundation (DFG STO1126/2-1, GL893/1-1, KA2679/3-1, and KFO 214).

\appendix
\section*{Appendix}
\section{Pseudocode for the proposed method}
A pseudocode of the proposed method is given in Algorithm~\ref{alg:estimation}. 

\begin{algorithm}[h]
\small
\caption{Proposed method for bolus arrival time estimation}
\label{alg:estimation}
\SetCommentSty{footnotesize}
\DontPrintSemicolon
\BlankLine
\KwIn{$c \in \R^{N}$: Given TCC}
\BlankLine
\textbf{Optional:} 
Range for spline orders $k_{\min{}}, k_{\max{}}$ (Default~values: 
$k_{\min{}} = 3, k_{\max{}} = 6$)\;
Coarse search space parameters 
$T_{\min{}}, \Delta T, T_{\max{}}, A_{\min{}}, 
\Delta A,A_{\max{}}$
(Default~values: $T_{\min{}} = t_{\min{}}$, $T_{\max{}} = t_{\max{}}$, $\Delta T = \Delta T/4$, $A_{\min{}} = 1$, $A_{\max{}} = 50$, $\Delta A = (A_{\max{}} - A_{\min{}})/25$)
\;
\BlankLine
\KwOut{$\tBAT^*$:  Estimated BAT}
\BlankLine
$S = \infty$\tcc*{Init cross validation score}

\For{$k = k_{\min{}},\ldots, k_{\max{}}$}{
        \BlankLine
        \tcc{Coarse search over discrete grid}
        $\Tc = \{T_{\min{}}, T_{\min{}} + \Delta T, T_{\min{}} + 2\Delta T, \ldots, T_{\max{}}\}$\;
        $\Ac = \{A_{\min{}}^{2k}, (A_{\min{}} + \Delta A)^{2k},  (A_{\min{}} + 2\Delta A)^{2k}, \ldots, A_{\max{}}^{2k} \}$\;
        $(t^\mathrm{cs}, \alpha^\mathrm{cs}) = \argmin_{t \in \Tc, \alpha \in \Ac} \log \GCV(c; t, \alpha, k)$
        \BlankLine
        \tcc{Fine search over continuous domain}
        Search minimizer $(t^\mathrm{fs}, \alpha^\mathrm{fs})$ of $\log \GCV(c; t, \alpha, k)$ 
        over the domain $[t_1, t_{N-k}] \times [\alpha_{\min{}}, \infty)$
        using Newtons method with starting point $(t^\mathrm{cs}, \alpha^\mathrm{cs})$\;
        \BlankLine
        \tcc{Check if (logarithmic) GCV score improved}
        $S^\mathrm{fs} = \log \GCV(c; t^\mathrm{fs}, \alpha^\mathrm{fs}, k)$\;
        \If{$S^\mathrm{fs} < S$ }{
            $S^\mathrm{fs} = S$\;
            $\tBAT^* = t^\mathrm{fs}$\;
        }
}
\end{algorithm}

\section{The method of Cheong \textit{et al.}}

The method of \citet{cheong2003automatic} is based on a parametric approximation model 
for the  concentration curve between the first time frame $t_1$
and the peak time denoted by $t_p.$
The model functions $u$ are assumed to be constant
before the BAT and a quadratic polynomial 
between the BAT and the peak time:
\begin{equation}\label{eq:cheong_model}
u(t_n) = 
\begin{cases}
\beta_0, & \text{if }t_1 \leq t_n \leq \tBAT, \\
\beta_0 + \beta_1 \cdot (t_n - \tBAT) + \beta_2 \cdot (t_n - \tBAT)^2, & \text{if } \tBAT < t_n \leq t_p.
\end{cases}
\end{equation}
Here, $n \in \{1, \ldots, p-2\}$ 
and $\beta_0, \beta_1, \beta_2$ are the model parameters.
We note that also a linear variant was proposed which is obtained by setting $\beta_2$ equal to zero. 
We found that in average imposing this restriction did not lead to improved estimates, thus we focus on the quadratic method. 
The method proceeds as follows.
For each BAT candidate $\tBAT \in \{ t_1, \ldots, t_\text{p}\},$
the optimal model parameters are computed by a least squares fit using non-negativity constraints.
The candidate that provides the least residual sum of squares is considered as the best estimate for the BAT.
Note that with this method
the estimated BAT is necessarily one 
of the sampled time points.

\section{Mathematical relation of the proposed method to the method of \citet{cheong2003automatic}}
As mentioned earlier, in the limit 
 $\alpha\to\infty,$ 
 the proposed spline model \eqref{eq:spline_model} reduces to least squares approximation by a polynomial of degree $k-1$ after the BAT.
 Hence, the unconstrained parametric model of  Cheong~\textit{et al.},
 i.e.~model \eqref{eq:cheong_model} without non-negativity constraints on $\beta$,
 appears as the limit case $\alpha\to\infty$  of the proposed non-parametric model of order $k = 3.$
 Furthermore, 
 minimizing the GCV for this case reduces to minimizing the residual sum of squares. This is because the denominator in \eqref{eq:GCV} is constant for polynomial approximation of fixed degree.
Also recall that Cheong's method 
limits the search space for the BAT to the sample time points
$t_1, \ldots, t_{p-2}.$
To summarize, the unconstrained variant of Cheong's method
can be formulated in terms 
of the proposed method \eqref{eq:GCV} as the limit $\alpha \to \infty$ of the optimization problem
$\argmin_{t \in \{t_1,\ldots, t_{p-2}\}} \GCV(c_{1:p}; t, \alpha, 3)$
where $c_{1:p} = (c_1, \ldots, c_p)$ denotes the signal until the time of maximum.
Therefore the proposed approach can be seen as generalization of 
the unconstrained variant of the method of \citet{cheong2003automatic}.

\renewcommand{\refname}{}
\setlength{\bibsep}{0pt plus 0.3ex}
\section*{References}
{\footnotesize
\vspace{-5ex}
\bibliographystyle{myagsm}
\setcitestyle{authoryear,open={(},close={)}}
\bibliography{refs}
}

\end{document}

%% file: Table_1.txt
\begin{tabular}{cccl@{\hskip 1ex}@{\hskip 3ex}r@{\hskip 1ex}r@{\hskip 0.5ex}r@{\hskip 3ex}r@{\hskip 1ex}r@{\hskip 0.5ex}rr@{\hskip 3ex}@{\hskip 3ex}r@{\hskip 1ex}r@{\hskip 0.5ex}r@{\hskip 3ex}r@{\hskip 1ex}r@{\hskip 0.5ex}rr} 
\toprule 
Animal & $M_{R}$ & $M_{L}$  & \multicolumn{1}{l}{Method} & \multicolumn{3}{c}{Right Tumor} & \multicolumn{3}{c}{Left Tumor} & \multicolumn{1}{l}{AIF} \\ 
\midrule 
1 & 7 & 31 & Cheong \textit{et al.} & 0.00 & [0.00, & 0.00] & 0.00 & [0.00, & 6.68] & 32.84 \\ 
  &  &  &  Cheong \textit{et al.} (adapted) & 25.37 & [25.37, & 25.37] & 25.37 & [25.37, & 25.37] & 32.84 \\ 
 & & &  Proposed & 35.47 & [34.88, & 36.04] & 35.41 & [34.60, & 36.30] & 32.82 \\[1ex] 
2 & 18 & 94 & Cheong \textit{et al.} & 0.00 & [0.00, & 4.93] & 0.00 & [0.00, & 6.57] & 34.33 \\ 
  &  &  &  Cheong \textit{et al.} (adapted) & 25.37 & [25.37, & 25.37] & 25.37 & [25.37, & 25.37] & 34.33 \\ 
 & & &  Proposed & 37.13 & [36.24, & 37.77] & 36.63 & [35.64, & 37.75] & 34.36 \\[1ex] 
3 & 32 & 26 & Cheong \textit{et al.} & 0.00 & [0.00, & 0.00] & 0.00 & [0.00, & 4.63] & 33.58 \\ 
  &  &  &  Cheong \textit{et al.} (adapted) & 25.37 & [25.37, & 25.37] & 25.37 & [25.37, & 25.37] & 33.58 \\ 
 & & &  Proposed & 36.90 & [35.00, & 38.28] & 36.81 & [35.94, & 37.86] & 33.81 \\[1ex] 
4 & 13 & 39 & Cheong \textit{et al.} & 0.00 & [0.00, & 19.93] & 0.00 & [0.00, & 11.42] & 32.09 \\ 
  &  &  &  Cheong \textit{et al.} (adapted) & 25.37 & [25.37, & 128.77] & 25.37 & [25.37, & 25.37] & 32.09 \\ 
 & & &  Proposed & 36.17 & [35.14, & 37.29] & 35.54 & [34.61, & 36.19] & 32.04 \\[1ex] 
5 & 13 & 7 & Cheong \textit{et al.} & 0.00 & [0.00, & 13.92] & 0.00 & [0.00, & 14.18] & 30.60 \\ 
  &  &  &  Cheong \textit{et al.} (adapted) & 25.37 & [25.37, & 25.37] & 25.37 & [25.37, & 25.37] & 30.60 \\ 
 & & &  Proposed & 33.32 & [33.04, & 34.10] & 33.47 & [33.22, & 33.86] & 30.50 \\ 
\bottomrule 
\end{tabular} 